# Decoding the Divide: Analyzing Disparities in Broadband Plans Offered by Major US ISPs

Udit Paul*, Vinothini Gunasekaran*, Jiamo Liu*, Tejas N. Narechania[§],
Arpit Gupta*, Elizabeth Belding*

University of California Santa Barbara*   University of California Berkeley[§]

## ABSTRACT

Digital equity in Internet access is often measured along three axes: availability, affordability, and adoption. Most prior work focuses on availability; the other two aspects have received little attention. In this paper, we study broadband affordability in the US. Specifically, we focus on the nature of broadband plans offered by major ISPs across the US. To this end, we develop a broadband plan querying tool (BQT) that obtains broadband plans (upload/download speed and price) offered by seven major ISPs for any street address in the US. We then use this tool to curate a dataset, querying broadband plans for over 837 k street addresses in thirty cities for seven major wireline broadband ISPs. We use a plan's carriage value, the Mbps of a user's traffic that an ISP carries for one dollar, to compare plans. Our analysis provides us with the following new insights: (1) ISP plans vary inter-city. Specifically, the fraction of census block groups that receive high and low carriage value plans varies widely by city; (2) ISP plans intra-city are spatially clustered, and the carriage value can vary as much as 600% within a city; (3) Cable-based ISPs offer up to 30% more carriage value to users when competing with fiber-based ISPs in a block group; and (4) Average income in a block group plays a critical role in dictating who gets a fiber deployment (i.e., a better carriage value) in the US. While we hope our tool, dataset, and analysis in their current form are helpful for policymakers at different levels (city, county, state), they are only a small step toward understanding digital equity. Based on our learnings, we conclude with recommendations to continue to advance our understanding of broadband affordability.

## 1 INTRODUCTION

The National Digital Inclusion Alliance (NDIA) defines digital equity as "a condition in which all individuals and communities have the information technology capacity needed for full participation in our society, democracy, and economy" [40]. As modern life has moved increasingly online, high-quality Internet access has become a key component of digital equity. The Covid-19 pandemic, and the post-pandemic "new normal" of remote interaction, have drastically changed the need for home Internet access; work-from-home, online/remote schooling, telemedicine, and other networked applications have become commonplace, to the point of nearly indispensable. As a result, individuals without home access to highly reliable, high-speed broadband are severely disadvantaged [34].

Having an in-depth understanding of digital equity is critical for policymakers at different levels (e.g., city, county, state, federal) to take corrective actions, such as offering subsidies [31], regulating rates [6] and funding access infrastructure [41]. Digital equity, especially in the context of Internet access, is often measured along three axes: availability, affordability, and adoption [46]. Many past efforts [28, 43, 44], including ones in our research community, have focused on measuring availability. Researchers have disaggregated availability into coverage and quality. Here, coverage answers whether broadband access is available in a geographical region, while quality answers questions related to access type (e.g., cable, fiber, DSL), and upload/download speed. Researchers and policymakers use publicly-available datasets, such as the FCC's Form 477 [30], Measuring Broadband America (MBA) [32], and Measurement Lab (M-Lab) speed test [38], as well as proprietary ones, such as Ookla's speed test [42], to characterize Internet connectivity. More recently, as part of the Broadband Equity, Access, and Deployment (BEAD) program, Congress directed the FCC to develop an accurate map of fixed broadband availability across the US. Though it is still a work in progress, when completed, the FCC National Broadband map will provide information regarding broadband availability (i.e., provider, access type, maximum upload/download speed) at the granularity of street addresses.

Though the existing datasets, including the most recent FCC National Broadband map, enable measuring availability, affordability has received less attention. To answer any question related to broadband affordability, extracting the "cost of broadband connectivity", i.e., the nature of the "deal" a user is getting, at fine-grained geographical granularity, is critical. Using the cost data, one can answer critical policy questions such as (1) what pricing policies do ISPs employ to users in different regions (i.e., neighborhoods, cities, states)?; (2) where, within a region, are different types of deals offered by ISPs?; (3) how does the (lack of) competition among ISPs affect broadband prices in a region?; and (4) how do socioeconomic and demographic factors affect broadband prices?

Most previous studies have either focused on manually querying ISP's websites [35, 39] or self-reporting from ISPs [33], and at best, they could scratch the surface of questions (1) and (2). A more recent study by a team of investigative journalists curated the broadband availability and cost data at street-level granularity for four major ISPs across 43 cities.[1] However, among other limitations, this study did not analyze the broadband plans for major cable-based ISPs (e.g., Cox), and thus, it could not fully answer questions (3) and (4).

Our goal is to curate a new dataset that enables a better understanding of broadband affordability in the US, addressing the limitations of prior related efforts. To this end, we present the design and implementation of a new *broadband plan querying tool (BQT)*. BQT takes a street-level address as input and returns the available broadband plans offered by major ISPs at that address. Here the plans entail the maximum upload speeds, download speeds, and corresponding prices in US dollars; typically, multiple plans are available to each residential address. BQT automates mimicking the behavior of a real user interacting with an ISP's website to query available broadband plans for a given street address. It addresses various challenges to offer a high hit rate, i.e., the number of street addresses it can successfully query for an ISP and the number of major ISPs it can query.

We use BQT to curate our broadband plans dataset while ensuring our data collection effort does not overwhelm ISP websites. Specifically, we collect and analyze plan data in thirty US cities with diverse populations, population density, and median income.[2] We identify seven major ISPs that reach 89% of the total census blocks in the US [12]. For each (ISP, city) pair, we sample a subset of residential addresses extracted from a dataset provided by Zillow [2]. We feed these addresses to BQT to curate the desired broadband affordability dataset.

We use this dataset to answer multiple policy questions about broadband affordability. Specifically, we use the metric *carriage value* to characterize broadband plans. A paper recently proposed this metric in the legal literature [39] that the White House referred to in announcing a new Executive Order [7] citing a call for arms to address the lack of competition among broadband service providers in the US. This metric quantifies the amount of user Internet traffic (Megabits) that an ISP carries per second, per dollar spent on a monthly broadband plan. For example, the carriage value for a broadband plan with a download speed of 100 Mbps at $50 is 2 Mbps/$. Intuitively, the higher the carriage value, the better the deal the user receives for their broadband subscription, and vice versa. We use this metric to study the quality of "deals" ISPs offer within and between different cities. From an end user's perspective, we explore how this metric varies across different ISPs active in a region, how the nature of the deal is affected by various demographic and socioeconomic factors and the state of competition among ISPs locally.

In summary, our work offers three major contributions:

**Broadband plan querying tool (Section 3).** We present the design and implementation of a broadband plan querying tool that reliably queries the websites of seven major ISPs, mimicking a real user, to extract the available broadband plans for a given street address.

**Broadband plans dataset (Section 4).** We present our methodology to curate a broadband plans dataset by querying 837 k unique addresses (1.2 M plans) across 30 cities (18 k census block groups) and seven major ISPs in the US.

**Characterization of broadband plans (Section 5).** We conduct a multi-dimensional analysis to study the intra- and inter-city distribution of broadband plans (i.e., carriage value) for each ISP and how these plans are affected by competition among ISPs and various demographic and socioeconomic factors. Our analysis offers the following key insights: (1) ISP plans vary by city, i.e., the fraction of census block groups that receive high (and low) carriage value plans are variable across cities. (2) ISP plans within a city are spatially clustered, and the carriage value can vary as much as 600% within a city. (3) Cable-based ISPs offer up to 30% greater carriage value to users when competing with fiber-based ISPs in a block group. (4) Average income in a block group plays a critical role in dictating who gets a fiber deployment (i.e., a better carriage value).

We view this work as an important step towards understanding broadband affordability in the US at scale. We note that broadband affordability is multifaceted, with numerous factors to consider. While our analysis provides valuable insight, it only scratches the surface of what policymakers must address when assessing broadband affordability. The evaluation of broadband affordability in a specific region or for a particular population may require consideration of additional factors beyond the scope of this paper. To enable other researchers and policymakers to advance our understanding of this critical topic, we will make our tool and a privacy-preserving version of our dataset **publicly available**. We conclude this study with recommendations for different stakeholders to further improve the understanding of broadband affordability.

---

[1] Our team provided technical assistance for this investigative reporting.
[2] Racial compositions of the cities we study are also diverse. However, we omit this aspect of our analysis at this time.

**Ethical concerns.** Please refer to Section 4.2 for a discussion of how we address ethical concerns regarding our data-collection tool and methodology.

## 2 BACKGROUND & MOTIVATION

**Broadband providers in the US.** Thousands of ISPs offer broadband connectivity, reaching approximately a hundred million residences in the US. Most of these ISPs operate locally and have a fairly small footprint [20, 25, 26]. This paper considers seven major ISPs, each serving at least one million residences. Together they reach 89% of the total census block groups in the US. We can divide these ISPs into two broad categories: fiber/DSL-based and cable-based providers. Our work, as well as [5], confirms that these ISPs either operate as a monopoly or duopoly, i.e., at max, only two major ISPs compete with each other in a census block group. Also, ISPs of the same type do not compete with each other: DSL/fiber-based ISPs do not compete with each other, and cable-based ISPs do not compete [5]. Moreover, in major cities, cable-based ISPs dominate in terms of coverage, i.e., they serve almost all the block groups [12]. In contrast, DSL/fiber providers serve a smaller fraction of block groups. Finally, in part because fiber deployments are relatively new and more expensive to deploy, DSL is often (though not always) offered in more block groups than fiber. Given these trends, cable-based ISPs operate in three distinct modes: *cable monopoly*, *cable-dsl duopoly*, and *cable-fiber duopoly*.

**Existing broadband availability datasets.** The FCC has recently launched a street-address level map of broadband availability [12]. This is an improvement over the previous iteration, based on provider input through Form 477 [30], that offered this information at the block-level granularity. This new map reports the maximum upload and download speeds and the access technology (e.g., fiber, cable) at street-level granularity and relies on self-reporting from ISPs. Previous efforts curated similar data by manually [19] or automatically [37] querying ISP web interfaces, also referred to as a broadband availability tool (BAT). Such third-party efforts enable auditing self-reported data from different ISPs [4, 24, 37].

These datasets improve our understanding of broadband availability, both in terms of coverage and quality. However, without any pricing information, it is not possible to characterize broadband affordability.

**Existing broadband plan datasets.** Many previous efforts have curated broadband plan datasets by manually querying ISPs' BATs. For example, the California Community Foundation and Digital Equity Los Angeles queried Spectrum's website to curate a list of broadband plans for 165 street addresses in Los Angeles county (California). The author in [39] manually compiled a dataset of 126 street addresses across seven states to obtain available plan information. While these studies highlight the disparity in broadband plans, small-scale datasets are, at best, suggestive of broader and more general trends

More recently, an online investigative platform, The Markup [13], extended the BAT client [37] to automate the extraction of broadband plans for four major ISPs in 43 US cities. Their study [14], which is the most closely related prior work to ours, finds significant variability in the download speed offered by major ISPs at different price points. For instance, the authors found that, for $55/month, AT&T offers 1000 times greater maximum download speed to some addresses in the same city; this phenomenon is referred to as "tier-flattening" [1]. The Markup study also finds that some major ISPs, such as AT&T and CenturyLink, provide lower speeds to more vulnerable populations, e.g., low-income and high-minority communities, than others. Based on this analysis, the authors highlight the importance of analyzing the cost of Internet service and download speed instead of download speed in isolation.

As discussed in Section 4.1, extending BAT clients to collect data for all major ISPs is non-trivial. Given these challenges, The Markup study does not include cable-based ISPs, which serve most of the US population [10]. Consequently, their dataset is not suited to explore the dynamics between cable and DSL/fiber providers nor the study of how competition between the two changes the nature of broadband plans in a region.

**Our approach.** In this work, we address the critical gaps of previous efforts by curating a comprehensive broadband plan dataset in terms of location and type of ISPs. First, we develop BQT to obtain plan information across 837 k street addresses for three major cable providers and four major DSL/Fiber providers. Our comprehensive dataset provides insights into the ISP plan structure across 30 cities around the country. Using this dataset, we can characterize how ISP plans change between cities, within a city, and in the presence of another ISP.

## 3 DEVELOP MEASUREMENT TOOL

Our goal is to develop a *robust* measurement tool that can *accurately* report the broadband plans offered by major ISPs for a given set of street-level addresses at *scale*. Rather than relying on user surveys [35] or self-reporting [33] from ISPs, we focus on directly querying ISP BATs. Minimizing disruptions to end users using BAT is our top priority while developing this tool. In essence, for a given list of input addresses, we want this tool to achieve a high hit rate, i.e., successfully extract broadband plans for as many input street addresses

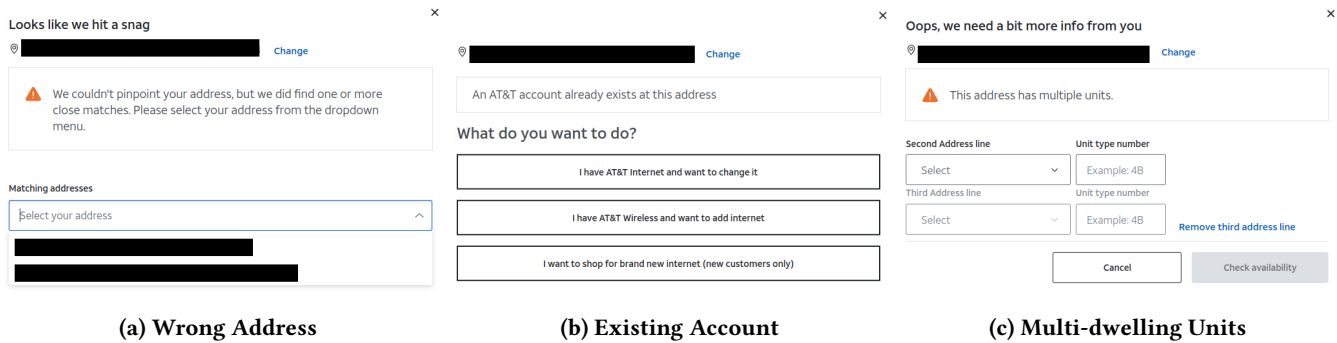

Figure 1: Illustration of different steps that BQT handles while querying ISP broadband plans through their BAT.

as possible, promptly, yet without disrupting the normal service offered by the ISP to end users.

### 3.1 Challenges

In theory, obtaining broadband plan information from an internet service provider's BAT should be straightforward. However, in practice, it is often complicated due to the quality of street address datasets. Most street address datasets are crowdsourced [22, 27], which can result in incomplete, incorrect, or ambiguous information. As a result, the querying process is a dynamic, multi-step process, where the information displayed at each step is based on the internal logic and state of each BAT, as well as the input provided by the user in the previous step. For instance, after the user enters a street address, the next web page may either show available broadband plans, indicate an incorrect input address, or inform the user that they are already a subscriber at that address. Additionally, ensuring that the tool can query all major ISPs is challenging because different ISPs use different formats and interfaces, such as drop-down menus or click buttons, to present this information and allow users to respond.

To illustrate, Figure 1 shows different steps that our tool needs to follow to extract the broadband plans. Here we use AT&T as an example but we can confirm that all other ISP BATs also follow these steps. In the first case as illustrated in Figure 1a, AT&T could not identify the input street address[3]. When faced with this scenario, the expected response for the end user is to access the drop-down menu that the BAT provides and further select an address from the set. As a next step, AT&T could indicate that an active customer already exists in this specific street address. Under this scenario, the BAT offers three distinct choices as shown in Figure 1b. If a user is already an AT&T's subscriber residing in that address, they can select the option of either changing a plan or adding a new plan. This would prompt the BAT to render an authentication form to ensure the user is an active subscriber. The third option that is provided applies to a new customer who are interested in viewing the set of AT&T's plan at that address. This step doesn't require any authentication. Finally, a particular address could be a multi-dwelling unit i.e. having an apartment/unit number that was not inputted during the initial stage. For that scenario, as demonstrated in Figure 1c, the BAT provides an option to select one of the possible apartments/units in that address.

---

[3]Note for privacy reasons, we have blurred the specific street address in this example.

### 3.2 Strawman: Extend Existing BAT Clients.

A potential solution to obtain broadband plan information is to enhance the BAT client proposed in previous research [37]. This tool was designed to query the binary availability of broadband service (i.e. (service/no service)) for a specific street address. The approach involves reverse-engineering the ISP's BAT by observing how it uses different RESTful APIs to extract the desired information, such as broadband availability. For example, the tool can observe that when a browser sends a request with a street address, it receives a response with an ID, and subsequent requests in the next step use this ID and, in some cases, a session cookie from the previous step. It then uses the Python `requests` library to directly send a series of requests to the ISP's RESTful APIs. Directly querying the APIs is scalable and can handle thousands of street addresses in parallel. In 2020, the authors used this tool to query approximately 35 million street addresses. Their data analysis revealed the limitations of the information provided by the FCC's Form 477 [30], reinforcing the need for such information to be made available at a street-level granularity as previously suggested by other research [29, 43].

**Limitations.** Since BAT has been successfully used to query millions of street addresses for all major ISPs, extending it to extract offered broadband plans seems like a natural choice. However, we observed that the proposed approach has several limitations that make it difficult to adapt to satisfy our goals. Specifically, since the publication of the previous work [37], ISPs have safeguarded their RESTful APIs from such direct querying[4]. For example, some ISPs have now started using dynamic cookies that append unique server-side parameters to each user session. Some BATs have started blocking queries from an IP address that uses the same cookie across multiple API requests. Dynamically generating a new cookie for each API request is non-trivial and is not supported by the original BAT client. Though it is possible to address some of these changes, the success (i.e., hit rate) of such an approach relies heavily on the nature of the safeguarding strategies employed by the ISP, which we have observed keeps evolving. Thus, we must develop a tool immune to such safeguarding strategies to achieve our stated goals.

## 3.3 BQT Approach

To decouple the querying process from ISPs' safeguarding strategies, our approach avoids directly querying their RESTful APIs. Instead, we use a popular web automation tool, Selenium, to mimic different end-user interactions for extracting the desired broadband plan information.

As a first step, we manually inspect the workflow for different ISP BATs. Each BAT employs a specific template to display the information for each step in the workflow. As part of this manual bootstrapping step, we enumerate all possible templates and identify unique patterns in their HTML content using regular expressions to help detect them at runtime.

The next step is to identify how to mimic a user's behavior using Selenium to advance successfully to the next step. This step is critical for ensuring a high hit rate for the proposed tool. Specifically, we handle different templates as described below.

Incorrect address. As we mentioned, street addresses are noisy due to inherent ambiguity between different identifiers. For example, for the same street address, some databases might use "Ave" instead of Avenue and "CT" or "Ct" instead of Court. Whenever there is a mismatch between the input street address and the one in ISP's database, it shows the "incorrect address" web page and often provides a list of one or more street addresses as suggestions. Given the prevalence of this issue, addressing it is critical to ensure a high hit rate for BQT. We address this issue by storing the list of suggested street addresses for offline analysis. We then apply string-matching over each suggested address in this list to find the one that best matches the input street address. As a sanity check, we ensure that the selected street addresses have the same zip code as our initially queried address. We then query the ISP's BAT to extract the broadband plan information.

---

[4]We do not assert that ISPs have changed their safeguarding strategies in response to previous data-collection efforts.

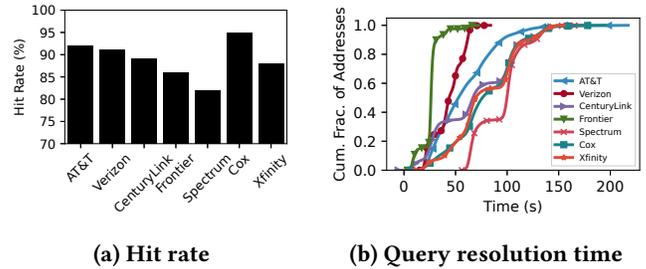

(a) Hit rate  (b) Query resolution time

Figure 2: BQT hit rate and distribution of query time distribution for different ISPs.

Multi-dwelling units. For addresses where a specific street address has multiple dwelling units (e.g., two or more apartments), the ISP BAT typically shows a "multi-dwelling unit" web page and suggests more refined street addresses (e.g., specific apartment numbers). Similar to previous work [37], we replace the input street address with a randomly selected address from this list. We then use this new address to query the ISP's BAT to extract the broadband plan information.

Existing customers. If the residents of an input street address are already subscribers, the ISP's BAT displays an "existing customer" web page and offers two options. The first option directs the user to their account, while the second allows a new user to query offered plans. Given our interest in extracting the available broadband plans, we chose the second option.

To avoid failures, we must ensure that all the Document Object Model elements for a step are successfully downloaded before applying any user action. The download times can vary across different templates and ISPs. For example, the step that displays available broadband plans after inputting the street address takes less than 30 seconds for AT&T but 60 seconds for Spectrum. Thus, we measure the download times for all possible templates and pause for this period (i.e., max observed download time) before applying the user action.

**Microbenchmarks.** The two crucial performance metrics of BQT are hit rate and query resolution time. The hit rate informs what fraction of total queried addresses we are able to successfully get a response for from a particular ISP BAT. As shown in Figure 2a, our hit rate for all ISPs exceeds 80%; we achieve the highest hit rate of 96 % for Cox and the lowest for Spectrum (82%). Such high hit rates across all ISPs ensure that BQT is able to extract plan information for the majority of the addresses. The query resolution time for a given street address is the amount of time it takes BQT to obtain a response from an ISP BAT. Figure 2b presents the distribution of query time for each ISP. The median time for Frontier query resolution is lowest, at 27 seconds, while it is highest, at 100 seconds, for Spectrum, despite no significant difference in the number of intermediate webpages rendered. In Section 4.1, we describe the methodology we adopt to make BQT more scalable.

**Limitations.** BQT has been specifically designed to work with the BATs offered by seven major ISPs. However, any changes made to the interfaces of these BATs by the ISPs, such as the addition of new drop-down forms, will require updating BQT. To ensure that BQT continues to function properly over time, we must monitor the BATs for all the supported ISPs and upgrade BQT as necessary to accommodate any changes. In the future, we plan to make BQT more modular, which will help minimize the effort required to adapt it to these changes.

## 4 CURATE BROADBAND PLANS DATASET

In this section, we describe the dataset we aggregate through BQT. We first describe our decision to query a subset of street addresses and ISPs to curate the desired broadband affordability dataset. We then describe how we selected the ISPs, cities, and street addresses for data collection (Section 4.1). Next, we discuss how we addressed different ethical concerns regarding our data-collection methodology (Section 4.2). Finally, we discuss the limitations of our dataset (Section 4.3).

### 4.1 Data Collection Methodology

In the US, approximately 7 ISPs serve around 90 million street addresses [12]. Through our data-collection agreement with Zillow [27], we have access to about 104 million "residential" US street addresses. Note that while this database does not represent every US address (it is comprised of addresses that had a transaction during a specific period), it encompasses a very large subset and provides an excellent starting point. Further, compared to alternative address datasets, such as the National Address Database (NAD) [22] offered by U.S. Department of Transportation, the Zillow dataset offers more complete coverage and is less noisy. Specifically, it includes nearly every county in the US, and USPS has validated the addresses as suitable for postal delivery [11]. Note that validation for postal delivery from USPS does not guarantee a perfect match with ISP's BAT; addresses can still be flagged as incorrect, incomplete, or ambiguous. However, it offers an excellent starting point.

**Diminishing returns.** In theory, we can use BQT to extract the available broadband plans for all ISPs that serve each street address in the Zillow dataset. However, we realized that curating such an extensive dataset has *diminishing returns* for the following reasons. First, our initial exploration of the collected data revealed that broadband plans are spatially clustered, i.e., plans for street addresses in the same neighborhood (i.e., a census block group) are similar. Second, Zillow's addresses are sparse in certain regions, especially rural and some sub-urban communities. Consequently, we cannot meaningfully characterize the nature of broadband plans in such areas. Third, with few exceptions, the smaller, local ISPs do not offer BATs that BQT could query. Moreover, these local ISPs have a significantly smaller service footprint in areas not as well covered in Zillow's street address data.

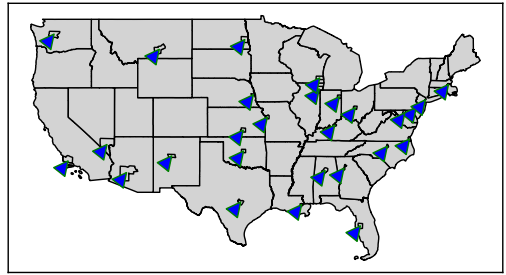

**Figure 3: Geographical location of the thirty US cities in our study.**

With this context in mind, we now describe how we selected a subset of ISP, cities, and street addresses for our study.

**ISP selection.** We focus on fixed broadband providers that offer queriable BATs and serve at least a million street addresses in Zillow's dataset[5]. After applying this filter, there are seven major ISPs: AT&T, Verizon, CenturyLink, Frontier, (Comcast) XFinity, (Charter) Spectrum, and Cox.[6] These ISPs together provide service to over 87% of the total US census blocks [12]. Among them, Xfinity, Spectrum, and Cox are cable-based providers, and AT&T, Verizon, CenturyLink, and Frontier offer DSL and fiber-based plans. Previous work reported that Comcast Xfinity's offerings are invariant to location [14]. Our analysis using the data collected using BQT from six major US cities confirmed these observations, and so we omit collecting data for this provider.

**City selection.** With a goal of wide geographic distribution, we examined cities with a range of population densities as well as diverse socioeconomic attributes (e.g., average income) that are well represented in the Zillow dataset. After applying this filter, we selected 30 major US cities in 27 states (see Figure 3). As shown in Table 2, these cities represent a broad spectrum of demographic and socioeconomic attributes. For example, the range of population densities

---

[5]We do not consider satellite-based ISPs as they offer low carriage value.
[6]We did not consider Windstream because it serves less than one million addresses [12].

varies from 1k to 42k, and the median yearly household income is from $31k to $101k.

**Street address selection.** Each city in the US is divided into census blocks, which are aggregated into census block groups. The US Census Bureau regards a census block group representing approximately 600-3000 people based on their demographic and socioeconomic characteristics. Thus to ensure that our sampling strategy mimics the socioeconomic composition of the city, we uniformly sample street addresses at the census block group level. Specifically, for each (ISP, city) pair, we identify the set of block groups covered by the ISP in a city. We randomly sample 10% of street addresses for each such block group. Note that this approach guarantees that we have at least thirty samples from each census block group for each ISP, which is crucial for ensuring the statistical significance of any metrics computed from these samples.

**Scaling data collection.** To gather the needed samples for our study, BQT needs to query 837 k street addresses. Hence we run multiple instances of BQT in parallel to scale the data collection. We use Docker containers to run these instances concurrently on a single local data-collection server. We can theoretically use as many containers as street addresses for different ISPs to expedite data collection. However, such an approach will overwhelm ISP BATs and degrade the user experience for its customers.

Though we cannot directly measure the experience for real users, we conducted an experiment where we measured ISP response time for 1, 50, 100, and 200 docker instances. We hypothesize that if running multiple dockers is affecting user experience, we should expect a statistically significant difference in ISP response time for different settings. However, we observe that the response time for any ISP did not change as we increased the number of docker instances. Based on this experiment, we can conclude that our choice of using up to 200 docker instances does not overwhelm ISP servers enough to disrupt the user experience. Nevertheless, we scale back and utilize 50-100 distinct containers for our data collection. Note that our choice of 200 instances is based on the intuition that an ISP should not get overwhelmed by such a small number. By no means it is an upper bound on how many docker containers we can run in parallel. However, exploring such an upper bound is disruptive and unethical.

To ensure that all our queries do not originate from a single non-residential IP address, we utilize a pool of residential IP addresses provided by Bright Initiative, the non-profit branch of Bright Data [18] (formerly known as Luminati). This organization offers free access to data scraping tools for nonprofits and academic organizations. Previous efforts [14, 37] have also used this service.

**Public release.** We will make this dataset publicly available to empower other researchers and policymakers to improve our communal understanding of broadband affordability in the US. To maintain the confidentiality of the street address information obtained from Zillow, we will take measures to protect its proprietary nature. This includes converting each street address within a census block group into a unique identifier using a hashing process before releasing our dataset.

### 4.2 Ethical Considerations

We use less than 100 docker instances at any given time to avoid overwhelming ISP BATs. We query ISP plans at the street address level and do not collect or analyze Personally Identifiable Information (PII). Our work does not involve human subjects research, and the private dataset provided by Zillow under the data use agreement does not reveal any individual's identity. Furthermore, the data gathered from the website does not include any PII. We do not have the means to identify residents, the selected broadband subscription tiers, or the actual performance received at any address.

### 4.3 Limitations

We now discuss a few limitations of our dataset and how to address them in the future.

**Staleness issues.** Our dataset only provides a single snapshot of broadband plans, which may change over time as ISPs update their infrastructure and pricing structures. We observe that many ISPs are actively deploying new fiber, and we expect their offered plans to change in the near future. We also noticed that ISPs occasionally offer discounts (i.e., higher *cv* plans), especially in areas where they compete with other major ISPs. Our dataset does not discriminate between normal and discounted offers and, thus, might not best reflect the most recent carriage for a subset of street addresses. We must repeat data collection at regular intervals (in order of months) to address these staleness issues.

**Limited coverage.** Currently, our dataset covers approximately 7.5% of total block groups in the US. Expanding it further requires better and more representative sources for street addresses. We currently use Zillow's data, which is biased toward high-density urban areas. We need a better representation of street addresses in semi-urban and rural areas. Though curating such datasets is challenging, recent efforts from the FCC to develop broadband availability maps at street address granularity demonstrate such an approach's feasibility.

**Veracity of reported plans.** There is no system or database to confirm the accuracy of the download speed and price data provided by the ISPs when querying a street address. However, as reported in [37], it is not in the interest of ISPs to report false or misleading information to potential customers, including poor performance or low-valued plans.

# 5 CHARACTERIZE BROADBAND PLANS

In this section, we will demonstrate how our broadband affordability dataset provides the means for various stakeholders to address crucial policy questions that previously were difficult to answer. To do so, we will first present an overview of the BQT dataset. We will then answer the following critical questions: ❶ Do the broadband plans, characterized by their carriage value, change by the city for different ISPs? ❷ Does the carriage value change within a city? If yes, which neighborhoods (identified by their census block groups) receive good and bad deals (high and low carriage values)? ❸ Does competition among ISPs impact the carriage value offered to the end users? If yes, is there a trend in which neighborhoods experience competition? ❹ Is the quality of available deals affected by demographic and socioeconomic factors? If yes, which population groups receive better or worse deals from the ISPs?

|  | Unique Plans | Download (Mbps) | Upload (Mbps) | Monthly Price ($) | cv |
|---|---|---|---|---|---|
| AT&T | 11 | 0.768-1000 | 0.768-1000 | 55-80 | 0.01-12.5 |
| Verizon | 4 | 3.1-1000 | 1-1000 | 50-100 | 0.4-11.1 |
| CenturyLink | 8 | 1.5-940 | 0.5-940 | 50-65 | 0.03-14.5 |
| Frontier | 2 | 0.2-2000 | 0.2-2000 | 50-100 | 0.0004-20.0 |
| Spectrum | 5 | 30-1000 | 5-35 | 20-70 | 11.1-14.3 |
| Cox | 6 | 100-1000 | 5-35 | 20-120 | 10.0-28.6 |
| Xfinity | 3 | 25-1200 | 5-35 | 20-80 | 3.8-15.0 |

Table 1: Overview of broadband plans offered by the seven major ISPs. The dashed line separates DSL/fiber-based providers from cable-based ones.

## 5.1 Dataset and Metrics

**Dataset overview.** Table 2 summarizes the number of street addresses and block groups we cover for each of the thirty cities. It also shows which of the seven major ISPs are active in each city and hence in our dataset. Overall, our dataset covers 837k distinct street addresses, representing 18k block groups (around 7.5% of total block groups in the US). None of the thirty cities are served by more than two major ISPs. Given the relatively low carriage value offered by local ISPs (if present), this trend indicates the presence of monopolies and duopolies in these cities [5].

Table 1 summarizes the available broadband plans from the seven major ISPs. The range of plans is more diverse for fiber/DSL-based providers than cable-based providers. The extremely low upload/download speeds (and related carriage values) are attributable to broadband plans via DSL.

**Calculating carriage values.** We use the carriage value to characterize a broadband plan offered by an ISP, and we curate this metric for all input street addresses. Since the entropy of available download speeds is greater than the upload speeds, we focus on download speed to calculate carriage value. While not shown, we verified that our results

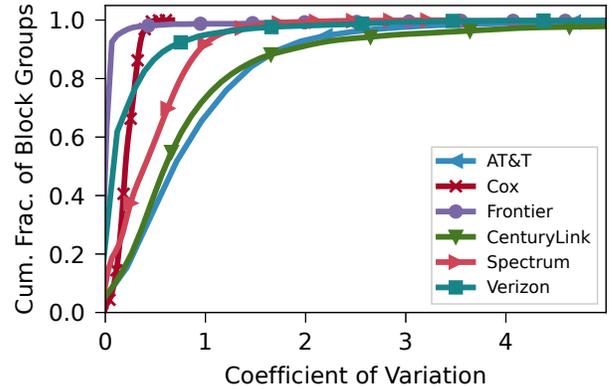

Figure 4: Distribution of coefficient of variation of carriage values in a block group for different ISPs.

are consistent if we use upload speed to determine carriage value.

Each ISP offers a fixed number of plans across all cities. For example, AT&T offers 11 different plans across the 14 cities it serves. However, an ISP only offers a subset of these plans at any given street address. For example, for a specific street address in New Orleans, AT&T offers three different plans: (1000 Mbps, $80/month), (500 Mbps, $65/month), and (300 Mbps, $55/month), which translates to carriage values of 12.5, 7.7, and 5.5, respectively. We report the best carriage value (*cv*) to characterize the available broadband plans at street-level granularity.

Reporting an aggregate metric at the granularity of the block group simplifies spatial analysis. Also, it ensures that our analysis is not biased by block groups with more street addresses in the Zillow dataset. To report the broadband plans at the block group granularity, we need to aggregate all *cv* for individual street addresses within a block group. However, such aggregations miss out on reporting the variability in broadband plans within the block group.

To best analyze broadband plans, we characterize the variability of individual values within a block group. Specifically, we compute the coefficient of variance (CoV), i.e., the ratio of the standard deviation to the mean of available plans within a block. Figure 4 shows how this metric varies across all the block groups for different ISPs. While we observe very low variability for most ISPs, there is a long tail for AT&T and CenturyLink. These providers offer DSL (very low CVs) and fiber plans (very high CVs) within the same block group. Given the low variability in carriage values within a block, we report the carriage value for a block group as the median of *cv* across all its street addresses.

**Comparing plans.** To compare an ISP's plans across different cities or the plans of two competing ISPs within a

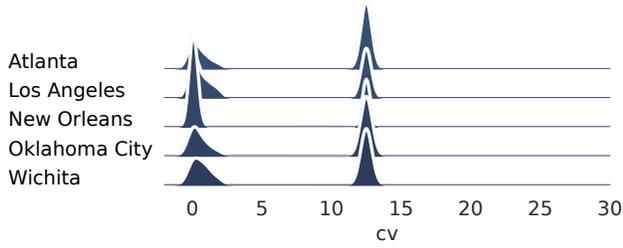

(a) AT&T (DSL/fiber)

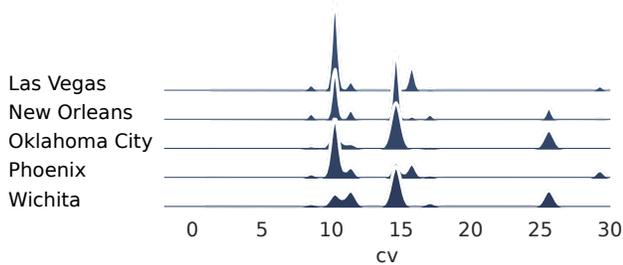

(b) Cox (cable)

Figure 5: Distribution of broadband plans in different cities for two major ISPs.

city, we need to quantify the differences in the plans. To this end, we represent the available plans from an ISP in a city using a plans vector of 30 dimensions, each representing a discrete carriage value.[7] We then quantify the differences using the L1 norm between the two vectors. The weight for each dimension is determined by the fraction of block groups in the city that receive that specific carriage value, and the ceil operator is used to discretize the carriage values. For example, Cox offers a carriage value of around 10.5 and 11.3 in 35%, and 12% block groups in New Orleans, 12% and 6%, block groups in Oklahoma City, and 4% and 21% block groups in Wichita. The L1 norm between New Orleans and Oklahoma City plans is 1.78 (different), between New Orleans and Wichita is 1.57 (different), and between Oklahoma City and Wichita is 0.36 (relatively similar).

## 5.2 Inter-City Broadband Plans

To answer ❶, we analyze the distribution of plans (at block group granularity). We only visualize one major provider from each DSL/Fiber (AT&T) and cable (Cox) category for brevity. To simplify the exposition, Figure 5 shows the distribution of carriage value for only five cities (out of 14 and 6, respectively) for each ISP.

---

[7]Note that the maximum carriage value we observed across all ISPs and cities is 28.6 (Table 1)

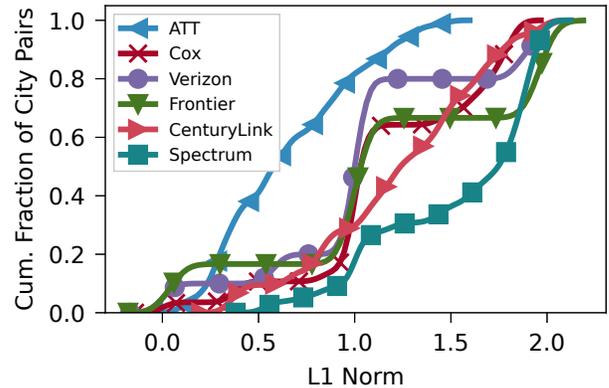

Figure 6: Distribution of difference in ISP plans across different city pairs. A higher L1 norm indicates more diverse offerings.

For AT&T, we observe two sets of peaks in broadband plans. The larger carriage value peak is attributable to fiber-based plans and the lower to DSL-based plans. Across different cities, we observe that the fraction of block groups receiving fiber plans differ. For example, in New Orleans, 32 % block groups receive fiber-based access, which is significantly smaller than 54 % and 57 % block groups in Wichita and Oklahoma City.

For Cox, we observe six different peaks, and the distribution of the carriage values across block groups varies significantly across different cities. For example, Cox offers *cv* of about 28 Mbps/$ to 7% of block groups in New Orleans. In contrast, Cox offers similar plans to 21% and 18% block groups in Oklahoma City and Wichita, respectively. On the other hand, 44%, 46%, and 50% of block groups in Wichita, New Orleans, and Oklahoma City receive *cv* of 14.6 Mbps/$.

To illustrate how this trend generalizes for other cities and ISPs, Figure 6 shows the distribution of L1 norm, i.e., the difference in available plans between all pairs of served cities for each ISP.[8] A low L1 norm indicates similarities in broadband plans and vice versa. We observe that DSL/fiber-based provider plans are less diverse across different cities than cable-based providers, with AT&T (most similar) and Spectrum (most diverse) at the extremes.

## 5.3 Intra-City Broadband Plans

To answer ❷, we zoom in on broadband plans within each city. At a high level, Figure 5 shows that ISPs offer disparate plans to users within a city. These differences can be as high as **600%** for DSL/fiber and **92%** for cable-based providers.
**Individual and composite plans.** To better understand broadband plans within a city, we zoom in on Cox and AT&T

---

[8]Recall Xfinity's offerings do not change within or between cities.

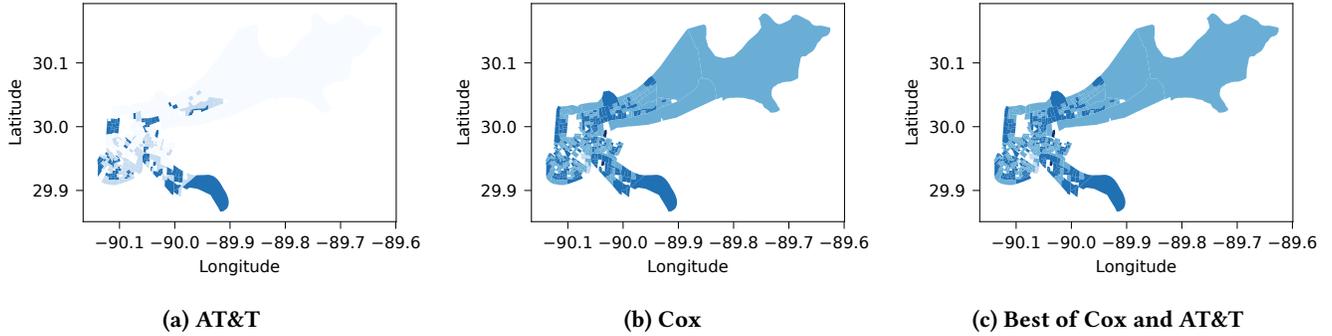

(a) AT&T  (b) Cox  (c) Best of Cox and AT&T

Figure 7: Spatial distribution of broadband plans in New Orleans. All three scenarios are spatially clustered.

in New Orleans, individually and as a pair (see Figure 7c). Comparing Figures 7a and 7b, we observe that Cox offers better coverage and higher carriage values than AT&T in most block groups.

Given its lower proliferation of high *cv* fiber plans, if one looks at the plans only from AT&T in this city, which was the case in one of the previous studies [14], we might get an impression that the nature of broadband plans is problematic for all New Orleans residents. Specifically, the broadband plans are sparse and highly variable (DSL vs. fiber), and most residents get the "worst" deal, i.e., low carriage values. However, the competing cable-based provider is the dominant ISP in the city, and its plans are not as extreme or sparse. Figure 7c shows that if we consider the plans from AT&T-Cox pair, i.e., when we report the highest carriage value from either of the two providers, we observe that the best carriage value is similar to that of the dominant cable-based ISP, i.e., Cox in this case. We make similar observations for other cities as well. In our dataset, we do not find a case where the DSL/fiber-based providers offer better coverage or higher average carriage values than the cable-based providers.

**Spatial clustering.** We (visually) observe that broadband plans are clustered, i.e., the likelihood that two contiguous block groups have similar available plans is high. To validate this visual understanding, we compute the spatial autocorrelation metric to characterize the extent of correlation in carriage values among nearby block groups. To this end, we employ Moran's I method [15]. This metric has been widely used in previous studies [36, 47] to understand the spatial distribution of a variable of interest (i.e., carriage value) within a geographic region (i.e., city). A positive value of Moran's I statistic means that similar carriage values tend to be found near each other, while a negative value means dissimilar values are found near each other, with zero indicating a complete lack of association of carriage values with locations.

We computed the Moran's I statistic for all the pairs of (ISP, city) to measure the spatial autocorrelation of broadband plans. Table 3 reports the median value across all cities for each ISP. The results show that, with the exception of Xfinity, the median value ranges between 0.3-0.5, indicating a high level of spatial clustering in broadband plans across ISPs within a city. Given that AT&T is a DSL/fiber-based provider, such clustering of its carriage value can be attributed to its fiber infrastructure deployment around the city.

Similar to the case for AT&T, the spatial clustering of plans for DSL/fiber providers is related to the nature of access technology. Neighborhoods with fiber deployments receive better carriage value and vice versa. However, since cable-based ISPs use the same technology across the city, spatial clustering in their plans is intriguing. In the next section, we explore whether this behavior is attributable to competition among ISPs.

### 5.4 Impact of Competition

To answer ❸, we explore whether the cable-based ISP's plans change when they operate as a monopoly vs. when they compete as a duopoly. We did not analyze DSL/fiber-based providers alone from the perspective of operating as both a monopoly and a duopoly because we did not observe this pattern in any of the thirty cities. To this end, we employ a statistical test to discern whether competition (or lack thereof) leads to a change in cable providers' carriage value. For every city with competition between cable and DSL/fiber providers, we run two one-tailed 2-sample Kolmogorov–Smirnov (KS) test [21].

Our null hypothesis ($H0$) is that there is no difference in the carriage value (*cv*) offered by a cable provider in locations where they operate as a cable monopoly compared to locations where they operate as a cable-dsl duopoly or cable-fiber duopoly. To test this hypothesis, we run one test for each of the following alternate hypotheses ($H$).

In the first one-tailed test, we propose $H1$, which states that the *cv* provided by the cable provider is greater for block groups in duopoly locations than those in cable monopoly locations. In the second test, we reverse the hypothesis from the previous test and propose $H2$, which states that cable providers provide better $cv$ for block groups in cable monopoly locations than those in each duopoly category. By conducting two tests per category, we can detect either scenario and provide robust statistical evidence of the impact of competition on cable offerings for different types of DSL/fiber-based offerings.

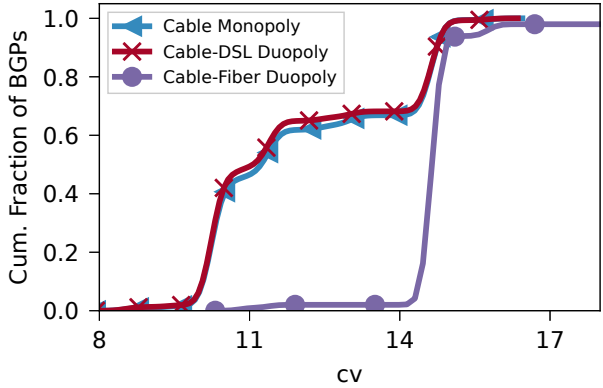

**Figure 8: Distribution of carriage value for Cox three operational modes in New Orleans.** To simplify exposition, we prune the long tail that is attributable to block groups receiving subsidized broadband access through the ACP plan [31].

If we achieve a p-value of less than 0.05, we reject the null hypothesis ($H0$) for the corresponding test and record the corresponding KS test statistic, denoted by the $D$ value. In the remainder of the section, we use New Orleans as a case study to explain our findings.

**Cable-DSL Duopoly:** In the first test, our $H1$ is Cox's $cv$ in cable monopoly block groups is lower than the cable-DSL duopoly block groups. Conversely, our $H2$ is Cox's $cv$ in cable monopoly block groups is higher than cable-dsl duopoly block groups. Figure 8 shows that Cox's offered $cv$ in the DSL duopoly block groups is similar to its $cv$ in monopoly block groups. This is further confirmed through the K-S test, where we fail to reject $H0$, which signifies there is no statistical difference in Cox's $cv$ distribution in block groups where it serves alone and block groups where it competes with AT&T's DSL offerings. The median $cv$ for both cases is 11.38 Mbps/$. We observe the same trend for other pairs of Cable-DSL Duopolies within cities in our dataset.

**Cable-Fiber Duopoly:** We posit a similar hypothesis for cable-fiber duopolies. Figure 8 shows the difference in Cox's $cv$ distributions between these block group types, which is further reinforced by the K-S test where we reject $H0$ with statistical significance in favor of $H1$ (with $D$ value of 0.65).

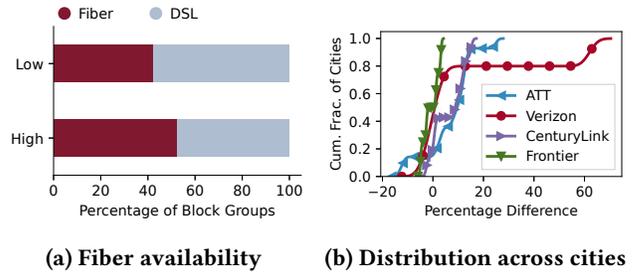

(a) Fiber availability  (b) Distribution across cities

**Figure 9: AT&T's DSL/fiber deployment.** In (a), the percentage of addresses served by the two technology types is broken down by income level in New Orleans. In (b), the overall distribution of the percentage difference in fiber deployment between high-income and low-income block groups across all cities is shown.

Contrarily, $H2$ cannot be accepted as the p-values exceed 0.05. This result points towards Cox increasing the $cv$ provided through its plans by lowering the price for the same download speed in block groups where it faces competition from AT&T's higher $cv$ fiber offerings. The median $cv$ from Cox in such addresses is 14.63 Mbps/$, 30% more than the monopoly and DSL block groups' median $cv$. For the remaining combinations of cable and DSL/fiber providers in other cities, we capture the same trend indicating differential pricing structures from cable providers in the presence of high $cv$ competition.

Our analysis in this section has demonstrated that cable providers tend to improve the carriage value offered through their plans in locations where fiber offerings are present. This places fiber-based offerings in a critical position because they tend to yield better broadband deals.

### 5.5 Influencing Socioeconomic Factors

In the prior sections, we established that low $cv$ is associated with DSL plans. In this section, we investigate whether there is a trend in which sociodemographic groups predominantly receive DSL plans, and therefore worse $cv$. This analysis will enable us to answer ❹. To do so, we compute the percentages of block groups within every city that receive DSL or fiber plans disaggregated by the block group level median household income. The American Community Survey (ACS) [9] publicly releases this information through a 5-year dataset. Although the demographic information for the 2020 census survey is available, it is known to have a significantly lower number of responses due to the COVID-19 pandemic [23]; hence we utilize the 2019 dataset. We merge our dataset with the ACS data to obtain the median household income of every census block group.

Concretely, we adopt a methodology similar to [14, 28] to group each city's census block group-level income into two distinct categories: low (below the city's median household income) and high (exceeding the city's median household income). For each income group class within a city, we calculate the percentage of block groups that have access to fiber-based plans. Subsequently, we compute the percentage difference in fiber deployment between the high and low-income groups of the block group.

Figure 9a presents the breakdown of the percentage of block groups that receive AT&T's DSL and fiber plans in the two income categories of block groups in New Orleans. 41% of the low-income census block groups receive AT&T's fiber plans while 57% of the high-income block groups in have fiber plans available. In 10 of the other 13 cities where we collected AT&T plan data, there is a greater presence of fiber-based plans for addresses in the high-income census block groups than the low-income block groups. As illustrated in 9b; a similar trend is also captured for CenturyLink and Verizon; Frontier is an outlier.

This result indicates greater fraction population in low-income block groups end up at the receiving end of low *cv* DSL plans from the fiber providers. Given that the lack of fiber also leads to lower *cv* from cable providers, the internet users of these block groups tend to get more bad deals overall compared to their counterparts in higher-income block groups. We also conducted a similar analysis for the demographic attributes of race and population density. The results for these variables did not produce similar trends across cities for all ISPs.

## 6 RELATED WORK

In [5], the authors analyzed the FCC Form 477 data and reported that close to 50 million people in the U.S. live in locations served by a single ISP, i.e. in an ISP monopoly. While not considering price/cost associated with internet access, several studies have sought to understand how internet quality itself varies between different locations and demographic variables. The Census Bureau produces an annual list of cities with the lowest Internet connectivity in the US, using data from the American Community Survey (ACS) One Year estimates [17]. However, this estimate does not take into account cost of access. The work conducted in [37] demonstrated that the FCC National Broadband Report significantly overestimates coverage and examined the digital divide in terms of the lack of coverage in rural and marginalized communities. In [3], the authors analyzed the relationship between income and download speed at the geographic granularity of U.S. zip codes. The work utilized income data, grouped into five income bins, obtained from 2017 tax returns filed with the Internal Revenue Service. The study demonstrated a positive correlation between zip code income and download speed. The authors of [28] analyzed publicly available data from Ookla [42], a popular speed test vendor, and found significant differences in key internet quality metrics between communities with different income levels. In [43], the authors utilized M-Lab [38] speed test data in California and found higher internet quality in urban and high-income areas.

Several studies have also examined how the cost of electricity varies across locations and demographic variables. The authors in [45] discovered that minority groups in various cities in the U.S. pay a disproportionate amount for electricity compared to other communities. Similar findings are also reported in [16].

## 7 CONCLUSION

In this work, we explore broadband affordability in the US. Specifically, we analyze the nature of broadband plans offered by sever major ISPs across thirty different cities in the US To aid this effort, we developed a new tool that extracts broadband plans offered by any of the seven major ISP for any street address. We use this tool to curate a dataset that reports broadband plans offered to 837 k street addresses, spanning 18 k census block groups and 30 cities. Our analysis sheds light on pricing strategies adopted by different ISPs, which have previously been opaque. Our results highlight the importance of competition. Specifically, it sheds light on how the ongoing fiber deployments benefit the end users. It also identifies the population groups reaping the benefits of competition and fiber deployments. We believe this effort is a small first step in the right direction to improve public understanding of broadband affordability in the US. We will make our tool and dataset publicly available to facilitate further research.

Drawing from our experiences, we recommend that regulators and policymakers take the following actions. (1) The FCC should consolidate the broadband availability maps [12] and urban rate survey [33] to ensure that the public has access to both availability and pricing information at the street address level. (2) Even if the FCC provides such data, third-party audits are essential to verify the accuracy of self-reported information from ISPs. However, existing street address datasets are private, sparse, and noisy, posing a challenge to such third-party efforts. Therefore, local governments (e.g., county) should put more effort into improving the quality and availability of street address datasets in their areas. (3) Policymakers should consider subsidizing fiber deployment efforts [8] or enforcing rate regulations [6] to help improve the carriage value for broadband plans in low-income block groups that often get ignored by major ISPs.

## REFERENCES

[1] 2018. TIER FLATTENING: AT&T and Verizon Home Customers Pay a High Price for Slow Internet. (2018). Retrieved 01/16/2023

[1] from https://www.digitalinclusion.org/wp-content/uploads/2018/07/NDIA-Tier-Flattening-July-2018.pdf

[2] 2018. Zillow's Transaction and Assessment Database (ZTRAX). (2018). Retrieved 02/14/2023 from https://www.zillow.com/research/ztrax/

[3] 2020. Decoding the digital divide. (2020). Retrieved 02/05/2023 from https://www.fastly.com/blog/digital-divide

[4] 2020. FCC Reports Broadband Unavailable to 21.3 Million Americans, BroadbandNow Study Indicates 42 Million Do Not Have Access. (2020). Retrieved 02/14/2023 from https://broadbandnow.com/research/fcc-underestimates-unserved-by-50-percent

[5] 2020. Profiles of Monopoly: Big Cable and Telecom. (2020). Retrieved 02/07/2023 from https://cdn.ilsr.org/wp-content/uploads/2020/08/2020_08_Profiles-of-Monopoly.pdf

[6] 2021. Assembly Bill A6259A. (2021). Retrieved 02/14/2023 from https://www.nysenate.gov/legislation/bills/2021/A6259#:~:text=Requires%20broadband%20providers%20to%20offer,if%20proper%20notice%20is%20given.

[7] 2021. FACT SHEET: Executive Order on Promoting Competition in the American Economy. (2021). Retrieved 01/16/2023 from https://www.whitehouse.gov/briefing-room/statements-releases/2021/07/09/fact-sheet-executive-order-on-promoting_competition_in-the_american-economy/

[8] 2021. Senate Bill No. 156. (2021). Retrieved 02/14/2023 from https://leginfo.legislature.ca.gov/faces/billTextClient.xhtml?bill_id=202120220SB156

[9] 2022. American Community Survey 5-Year Data (2009-2021). (2022). Retrieved 02/07/2023 from https://www.census.gov/data/developers/data-sets/acs-5year.html

[10] 2022. Cable Internet in the USA. (2022). Retrieved 01/16/2023 from https://broadbandnow.com/Cable

[11] 2022. DPV | PostalPro. (2022). Retrieved 02/07/2023 from https://postalpro.usps.com/address-quality/dpv

[12] 2022. FCC National Broadband Map. (2022). Retrieved 01/15/2023 from https://broadbandmap.fcc.gov/home

[13] 2022. The Markup. (2022). Retrieved 01/16/2023 from https://themarkup.org/

[14] 2022. The Markup. (2022). Retrieved 01/16/2023 from https://themarkup.org/still-loading/2022/10/19/dollars-to-megabits-you-may-be-paying-400-times-as-much-as-your\protect\discretionary{\char\hyphenchar\font}{}{}neighbor-for-internet-service

[15] 2022. Moran's I. (2022). Retrieved 02/07/2023 from https://en.wikipedia.org/wiki/Moran%27s_I

[16] 2022. Race and energy poverty: Evidence from African-American households. *Energy Economics* 108 (2022), 105908.

[17] 2023. American Community Survey 1-Year Data (2005-2021). (2023). Retrieved 02/05/2023 from https://www.census.gov/data/developers/data-sets/acs-1year.html

[18] 2023. bright data. (2023). Retrieved 02/07/2023 from https://brightdata.com/

[19] 2023. BroadbandNow. (2023). Retrieved 02/14/2023 from https://broadbandnow.com/

[20] 2023. Consolidated Communications. (2023). Retrieved 02/14/2023 from https://www.consolidated.com/

[21] 2023. Kolmogorov–Smirnov test. (2023). Retrieved 02/07/2023 from https://en.wikipedia.org/wiki/Kolmogorov%E2%80%93Smirnov_test

[22] 2023. National Address Database. (2023). Retrieved 02/14/2023 from https://www.transportation.gov/gis/national-address-database

[23] 2023. Sample Size. (2023). Retrieved 02/07/2023 from https://www.census.gov/acs/www/methodology/sample-size-and-data-quality/sample-size/index.php

[24] 2023. Senators Fear 'Deeply Flawed' FCC Broadband Map Could Screw Them Out of Millions in Federal Funds. (2023). Retrieved 01/16/2023 from https://gizmodo.com/senators-fcc-broadband-map-deeply-flawed-federal-fund-1849975157

[25] 2023. Utah Telecommunication Open Infrastructure Agency. (2023). Retrieved 02/14/2023 from https://en.wikipedia.org/wiki/Utah_Telecommunication_Open_Infrastructure_Agency

[26] 2023. WOW! Logo. (2023). Retrieved 02/14/2023 from https://www.wowway.com/

[27] 2023. Zillow. (2023). Retrieved 02/07/2023 from https://www.zillow.com/

[28] Francesco Bronzino, Nick Feamster, Shinan Liu, James Saxon, and Paul Schmitt. 2021. Mapping the Digital Divide: Before, During, and After COVID-19. In *Conference on Communications, Information, and Internet Policy (TPRC)*. https://papers.ssrn.com/sol3/papers.cfm?abstract_id=3786158

[29] Cooperative Network Services. 2023. RDOF and flawed 477 reporting. (2023). Retrieved 02/14/2023 from https://www.cooperative-networks.com/rdof-477-reporting/

[30] Federal Communication Commission. [n. d.]. ([n. d.]).

[31] Federal Communication Commission. 2023. Affordable Connectivity Program. (2023). Retrieved 02/07/2023 from https://www.fcc.gov/acp

[32] Federal Communication Commission. 2023. Measuring Broadband America. (2023). Retrieved 02/07/2023 from https://www.fcc.gov/general/measuring-broadband-america

[33] Federal Communication Commission. 2023. Urban Rate Survey Data & Resources. (2023). Retrieved 02/07/2023 from https://www.fcc.gov/economics-analytics/industry-analysis-division/urban-rate-survey-data-resources

[34] Amanda Holpuch. 13. US's Digital Divide 'is going to kill people' as COVID-19 exposes Inequalities. https://www.theguardian.com/world/2020/apr/13/coronavirus-covid-19-exposes-cracks-us-digital-divide. (Apr 13). (Accessed on 05/10/2020).

[35] LA Times 2022. Broadband internet isn't equally available to L.A. County's low-income residents, report says. (2022). Retrieved 01/16/2023 from https://www.latimes.com/business/story/2022-10-13/broadband-internet-not-equally-available-to-la\-county-low-income-residents-report-says

[36] Ossola Alessandro Locke, Dexter Henry, Emily Minor, and Brenda B. Lin. 2022. Spatial contagion structures urban vegetation from parcel to landscape. *People and Nature.* (2022).

[37] David Major, Ross Teixeira, and Jonathan Mayer. 2020. No WAN's Land: Mapping U.S. Broadband Coverage with Millions of Address Queries to ISPs. In *Proceedings of the ACM Internet Measurement Conference (IMC '20)*. 393–419.

[38] MLAB 2023. MLAB Test Your Speed. (2023). Retrieved 02/06/2023 from https://speed.measurementlab.net/#/

[39] Tejas N. Narechania. 2021. Convergence and a Case for Broadband Rate Regulation. *Berkeley Technology Law Journal* (2021).

[40] National Digital Inclusion Alliance. 2021. Definitions – National Digital Inclusion Alliance. (2021). Retrieved 02/14/2023 from https://www.digitalinclusion.org/definitions/

[41] National Telecommunications and Information Administration. 2022. Broadband Equity Access and Deployment (BEAD) Program. (2022). Retrieved 01/14/2023 from https://grants.ntia.gov/grantsPortal/s/funding-program/a0g3d00000018ObAAI/broadband-equity-access-and-deployment-bead-program

[42] Ookla 2023. SPEEDTEST. (2023). Retrieved 02/06/2023 from https://www.speedtest.net/

[43] Udit Paul, Jiamo Liu, David Farias-llerenas, Vivek Adarsh, Arpit Gupta, and Elizabeth Belding. 2022. Characterizing Internet Access and

Quality Inequities in California M-Lab Measurements. In *ACM SIGCAS/SIGCHI Conference on Computing and Sustainable Societies (COMPASS)*.

[44] Udit Paul, Jiamo Liu, Mengyang Gu, Arpit Gupta, and Elizabeth Belding. 2022. The Importance of Contextualization of Crowdsourced Active Speed Test Measurements *(IMC '22)*.

[45] Eric Scheier and Noah Kittner. 2022. A measurement strategy to address disparities across household energy burdens. *Nat Commun 13* 288 (2022).

[46] Geoffrey Starks. 2022. Availability, Adoption, and Access: The Three Pillars of Broadband Equity. (2022). Retrieved 02/07/2023 from http://soba.iamempowered.com/availability-adoption-and-access-three-pillars-broadband-equity

[47] Bell Nathaniel Zahnd, Whitney E. and Annie E. Larson. 2021. Geographic, racial/ethnic, and socioeconomic inequities in broadband access. *The Journal of Rural Health* (2021).

# APPENDIX

Table 2: Dataset coverage. The major ISPs are listed in the following order: (1) ATT, (2) Verizon, (3) CenturyLink, (4) Frontier, (5) Spectrum, (6) Cox, and (7) Xfinity.

|  | Block Groups | Street Addresses (k) | Population Density (k) | Median Income (k) | Major ISPs |  |  |  |  |  |  |
|---|---|---|---|---|---|---|---|---|---|---|---|
|  |  |  |  |  | 1 | 2 | 3 | 4 | 5 | 6 | 7 |
| Albuquerque, NM | 387 | 14 | 1.8 | 53 |  |  | • |  |  |  |  |
| Atlanta, GA | 389 | 12 | 1.2 | 65 | • |  |  |  |  |  | • |
| Austin, TX | 487 | 25 | 1.7 | 74 | • |  |  |  | • |  |  |
| Baltimore, MD | 1188 | 42 | 1.7 | 81 |  | • |  |  |  |  | • |
| Billings, MT | 98 | 3 | 1.1 | 61 |  |  | • |  | • |  |  |
| Birmingham, AL | 354 | 24 | 716 | 47 | • |  |  |  | • |  |  |
| Boston, MA | 37 3 | 17 | 8.4 | 72 |  | • |  |  |  |  | • |
| Charlotte, NC | 472 | 21 | 2 | 73 | • |  |  |  | • |  |  |
| Chicago, IL | 1933 | 86 | 3.8 | 64 | • |  |  |  |  |  | • |
| Cleveland, OH | 754 | 35 | 4.8 | 31 | • |  |  |  | • |  |  |
| Columbus, OH | 662 | 20 | 1.9 | 58 | • |  |  |  | • |  |  |
| Durham, NC | 138 | 5 | 1 | 59 |  |  |  | • | • |  |  |
| Fargo, ND | 67 | 5 | 1.5 | 62 |  |  | • |  |  |  |  |
| Fort Wayne, IN | 209 | 11 | 0.9 | 54 |  |  |  | • |  |  | • |
| Kansas City, MO | 305 | 15 | 1.2 | 51 | • |  |  |  | • |  |  |
| Los Angeles, CA | 1787 | 90 | 8.5 | 67 | • |  |  |  | • |  |  |
| Las Vegas, NV | 881 | 38 | 1 | 65 |  |  | • |  |  | • |  |
| Louisville, KY | 505 | 41 | 1.6 | 56 | • |  |  |  | • |  |  |
| Milwaukee, WI | 560 | 27 | 2.9 | 50 | • |  |  |  | • |  |  |
| New Orleans, LA | 439 | 67 | 2.9 | 41 | • |  |  |  |  | • |  |
| New York City, NY | 1567 | 51 | 41.7 | 96 |  | • |  |  | • |  |  |
| Oklahoma City, OH | 493 | 20 | 1.3 | 50 | • |  |  |  |  | • |  |
| Omaha, NE | 455 | 28 | 1.7 | 62 |  |  | • |  |  | • |  |
| Philadelphia, PA | 981 | 32 | 8 | 46 |  | • |  |  |  |  | • |
| Phoenix, AZ | 802 | 32 | 1.9 | 64 |  |  | • |  |  | • |  |
| Santa Barbara, CA | 211 | 6 | 2 | 79 |  |  |  |  | • | • |  |
| Seattle, WA | 634 | 28 | 2.1 | 101 |  | • |  |  |  |  |  |
| Tampa, FL | 536 | 25 | 1.5 | 57 |  |  |  | • | • |  |  |
| Virginia Beach City, VA | 112 | 4 | 1.8 | 80 |  | • |  |  |  | • |  |
| Wichita, KS | 304 | 13 | 1.3 | 50 | • |  |  |  |  | • |  |
| **Total** | **18k** | **837** |  |  | **14** | **5** | **7** | **4** | **13** | **8** | **6** |

## Individual ISPs

| 1 | 2 | 3 | 4 | 5 | 6 | 7 |
|---|---|---|---|---|---|---|
| 0.34 | 0.52 | 0.33 | 0.45 | 0.23 | 0.35 | 0 |

## ISP Pairs

| 1-5 | 1-6 | 3-5 | 3-6 | 4-5 | 2-5 | 2-6 | 1-7 | 2-7 | 3-7 |
|---|---|---|---|---|---|---|---|---|---|
| 0.23 | 0.35 | 0.23 | 0.35 | 0.23 | 0.23 | 0.35 | 0 | 0 | 0 |

Table 3: Statistical evidence for spatial clustering. We report the median of Moran I statistics across all cities.